\documentclass[aps,preprint,tightenlines]{revtex4}%
\usepackage{amsmath}
\usepackage{graphicx}%
\usepackage{amsfonts}%
\usepackage{amssymb}

\begin{document}

\title{Cyclotron Resonance of an Interacting Polaron Gas in a Quantum Well.
Magneto-Plasmon-Phonon Mixing}
\author{S. N. Klimin$^{\ast}$ and J. T. Devreese$^{\ast\ast}$}
\affiliation{Theoretische Fysica van de Vaste Stoffen, Universiteit Antwerpen (U.I.A.),
B-2610 Antwerpen, Belgium}
\date{August 26, 2003}

\begin{abstract}
Cyclotron-resonance (CR) spectra of a gas of interacting polarons confined in
a GaAs/AlAs quantum well are theoretically investigated taking into account
the magneto-plasmon-phonon mixing and band nonparabolicity. Contributions of
different magneto-plasmon-phonon modes to the total magneto-polaron coupling
strength are investigated as a function of the electron density. It is
confirmed theoretically, that the resonant magneto-polaron coupling in a
high-density GaAs/AlAs quantum well occurs near the GaAs TO-phonon frequency
rather than near the GaAs LO-phonon frequency. Calculated CR spectra are in
agreement with recent experimental data.

\end{abstract}
\maketitle

\section{Introduction}

The electron-LO-phonon interaction plays an important role in the optical
properties of polar semiconductors and ionic crystals (see, for review,
Refs.~\cite{KW,Appel,D1972,Polarons,Calvani2001}). The application of an
external magnetic field adds new features to the optical absorption spectra of
these materials. When the cyclotron frequency $\omega_{c}$ is close to the
LO-phonon frequency, resonant magneto-polaron coupling (i. e., anticrossing of
zero-phonon and, e. g., one-phonon states of the polaron system) occurs
\cite{Larsen1964}. Experiments on the cyclotron resonance (CR) in bulk
\cite{McCombe68} and quasi-2D \cite{Chang88,Cheng91,Peeters92} systems give
clear evidence of this resonant magneto-polaron coupling. However, CR
measurements on semiconductor quantum wells with high electron density
\cite{Ziesmann1987,Poulter2001} reveal that the anticrossing occurs near the
TO-phonon frequency rather than near the LO-phonon frequency. In
Ref.~\cite{Poulter2001}, on the basis of a model dielectric function for a 3D
medium, it was suggested that in a GaAs/AlAs quantum well with high electron
density, there may exist longitudinal modes, with energies close to the
TO-phonon frequency, which are active in CR.

Another consequence of high electron densities combined with the
nonparabolicity of the conduction band is a splitting of the CR lines
\cite{McD1989,Hu1995,Hu1995-2,Manger2001,Bychkov2002}. It was shown in
Ref.~\cite{Manger2001}, that the one-particle approximation fails to interpret
the oscillator strengths of the split CR lines.

Many-polaron CR in a 2D polaron gas was investigated theoretically in
Ref.~\cite{DW87} using the memory-function technique \cite{Mori1965}. In
Ref.~\cite{Peeters92}, this technique was applied to a polaron gas in a
quantum well. The CR spectra of polarons in a quantum well were calculated in
Ref.~\cite{HPD93} and applied in Refs.~\cite{Wang97,L} taking into account the
electron-phonon interaction for both bulk-like and interface phonon modes. To
quantitatively explain the CR data for a high-density polaron gas, as observed
in Ref.~\cite{Poulter2001}, it is necessary to take into account many-body effects.

In the present paper, the CR spectra of an arbitrary-density polaron gas in a
quantum well are calculated using the many-body memory-function technique
\cite{Mori1965}. The approach developed in Refs.~\cite{DW87,HPD93} is extended
here to a system of electrons confined in a quantum well and interacting with
magneto-plasmon-phonon modes. Within this method, we take into account the
electron-electron interaction, the static and dynamic screening of the
electron-phonon interaction \cite{DW87}, the magneto-plasmon-phonon mixing
\cite{Wendler90}, the band nonparabolicity and the phonon spectra specific for
a quantum well.

The paper is organized as follows. In Sec. II, the optical conductivity for a
polaron gas in a quantum well is derived. In Sec. III, we discuss the
calculated CR spectra and compare them with experimental data for a GaAs/AlAs
quantum well. Sec. IV contains the conclusions.

\section{Theoretical approach}

\subsection{Hamiltonian}

We consider a finite-barrier quantum well of width $d.$ The quantum well
(medium 1) with high-frequency dielectric constant $\varepsilon_{1}$ is placed
into a matrix (medium 2) with high-frequency dielectric constant
$\varepsilon_{2}.$ Both these media are supposed to be polar. An external
magnetic field $\mathbf{B}$ is applied parallel to the $z$-axis. The symmetric
gauge is chosen for the vector potential of the magnetic field,%
\begin{equation}
\mathbf{A}\left(  \mathbf{r}\right)  =\frac{1}{2}\left[  \mathbf{r}%
\times\mathbf{B}\right]  . \label{A}%
\end{equation}

The Hamiltonian describing the system, which consists of electrons interacting
with phonons and with each other, is
\begin{align}
H  &  =\sum_{nlm\sigma}E_{nl\sigma}a_{nlm\sigma}^{+}a_{nlm\sigma}%
+\sum_{\lambda,\mathbf{q}}\hbar\omega_{\lambda,\mathbf{q}}b_{\lambda
,\mathbf{q}}^{+}b_{\lambda,\mathbf{q}}\nonumber\\
&  +\frac{1}{\sqrt{S}}\sum_{\lambda,\mathbf{q}}\sum_{n,k}\left(
\gamma_{\lambda,\mathbf{q}}\right)  _{nk}\rho_{nk}\left(  \mathbf{q}\right)
\left(  b_{\lambda,\mathbf{q}}+b_{\lambda,-\mathbf{q}}^{+}\right) \nonumber\\
&  +\frac{1}{2S}\sum_{\mathbf{q}}\sum_{n_{1}k_{1}n_{2}k_{2}}V_{C}\left(
k_{1},n_{1};n_{2},k_{2}|\mathbf{q}\right)  \mathcal{N}\left[  \rho_{k_{1}%
n_{1}}^{+}\left(  \mathbf{q}\right)  \rho_{n_{2}k_{2}}\left(  \mathbf{q}%
\right)  \right]  , \label{H}%
\end{align}
where $a_{nlm\sigma}^{+}$ $\left(  a_{nlm\sigma}\right)  $ is a creation
(annihilation) operator for an electron in the one-particle state with energy
$E_{nl\sigma}$ and with the wave function%
\begin{align}
\Psi_{nlm}\left(  \rho,\varphi,z\right)   &  =\psi_{n}\left(  z\right)
\phi_{lm}\left(  \rho,\varphi\right)  ,\nonumber\\
\phi_{lm}\left(  \rho,\varphi\right)   &  =\frac{1}{\sqrt{2\pi}}e^{im\varphi
}\Phi_{lm}\left(  \rho\right)  . \label{States}%
\end{align}
Here, the function $\psi_{n}\left(  z\right)  $ corresponds to the $n$-th
size-quantized subband for motion along the $z$-axis, while $\phi_{lm}\left(
\rho,\varphi\right)  $ characterizes the ``in-plane'' motion of an electron
with a definite $z$-projection $m$ of its angular momentum, $l$ is the Landau
level quantum number.

The phonon frequencies $\omega_{\lambda,\mathbf{q}}$ and the amplitudes of the
electron-phonon interaction $\gamma_{\lambda,\mathbf{q}}$ are explicitly
derived in Refs.~\cite{HPD93,HPD90}. The index $\lambda$ refers to the phonon
branches, $\mathbf{q}$ is the phonon two-dimensional wave vector,
$b_{\lambda,\mathbf{q}}^{+}$ ($b_{\lambda,\mathbf{q}}$) is a phonon creation
(annihilation) operator. $\rho_{nk}\left(  \mathbf{q}\right)  $ is the
electron density operator%
\begin{equation}
\rho_{nk}\left(  \mathbf{q}\right)  =\sum_{lml^{\prime}m^{\prime}\sigma
}\left(  e^{i\mathbf{q\cdot r}_{\parallel}}\right)  _{lm,l^{\prime}m^{\prime}%
}\hat{a}_{nlm\sigma}^{+}\hat{a}_{kl^{\prime}m^{\prime}\sigma} \label{density}%
\end{equation}
with the matrix element%
\begin{equation}
\left(  e^{i\mathbf{q\cdot r}_{\parallel}}\right)  _{lm,l^{\prime}m^{\prime}%
}=\int_{0}^{\infty}\rho d\rho\int_{0}^{2\pi}d\varphi e^{i\mathbf{q\cdot
r}_{\parallel}}\phi_{lm}^{\ast}\left(  \rho,\varphi\right)  \phi_{l^{\prime
}m^{\prime}}\left(  \rho,\varphi\right)  . \label{Mel}%
\end{equation}
In the Hamiltonian (\ref{H}), $V_{C}\left(  k_{1},n_{1};n_{2},k_{2}%
|\mathbf{q}\right)  $ is the matrix element of the electron-electron
interaction potential%
\begin{equation}
V_{C}\left(  k_{1},n_{1};n_{2},k_{2}|\mathbf{q}\right)  =\int_{-\infty
}^{\infty}dz\int_{-\infty}^{\infty}dz^{\prime}\tilde{V}_{C}\left(
\mathbf{q};z,z^{\prime}\right)  \psi_{k_{1}}^{\ast}\left(  z\right)
\psi_{n_{1}}\left(  z\right)  \psi_{n_{2}}^{\ast}\left(  z^{\prime}\right)
\psi_{k_{2}}\left(  z^{\prime}\right)  , \label{VC}%
\end{equation}
$S$ is the area of the quantum well in the $xy$-plane, $\mathcal{N}\left[
\ldots\right]  $ is a normal product of operators. The two-dimensional Fourier
amplitude $\tilde{V}_{C}\left(  \mathbf{q};z,z^{\prime}\right)  $ of the
electron-electron interaction potential, specific for the quantum well,
differs from the corresponding expression for the Coulomb potential in bulk
owing to the presence of the electrostatic image forces, which appear as long
as $\varepsilon_{1}\neq\varepsilon_{2}$ (see, e.~g.,
Refs.~\cite{PFB88,PFB90,Mon}).

Within the local parabolic band approximation \cite{Larsen1987,Wu1989},
nonparabolicity is considered as a small perturbation. Consequently, the wave
function $\Phi_{lm}\left(  \rho\right)  $ (\ref{States}) for an electron in a
magnetic field takes the form%
\begin{align}
\Phi_{lm}\left(  \rho\right)   &  =\sqrt{\frac{2K!}{\left(  \left|  m\right|
+K\right)  !}}\left(  \frac{m_{b}\omega_{c}}{2\hbar}\right)  ^{\frac{\left|
m\right|  +1}{2}}\rho^{\left|  m\right|  }\exp\left(  -\frac{m_{b}\omega_{c}%
}{4\hbar}\rho^{2}\right)  L_{K}^{\left(  \left|  m\right|  \right)  }\left(
\frac{m_{b}\omega_{c}}{2\hbar}\rho^{2}\right)  ,\label{States2}\\
K  &  =l-\frac{m+\left|  m\right|  }{2},\quad m=-\infty,\ldots,l\nonumber
\end{align}
with the Laguerre polynomial $L_{K}^{\left(  a\right)  }\left(  z\right)  $.
Here, the cyclotron frequency $\omega_{c}=eB/\left(  m_{b}c\right)  $
corresponds to the mass $m_{b}$ given by \cite{HPD93}
\begin{equation}
m_{b}=\frac{m_{b1}m_{b2}}{P_{w}m_{b2}+P_{b}m_{b1}}, \label{mb}%
\end{equation}
where $P_{w}$ ($P_{b}$) is the probability to find the electron inside
(outside) the quantum well, and $m_{bi}$ ($i=1,2$) is the electron band mass
at the bottom of the conduction band in the $i$-th medium. The energies of the
spin-dependent electron states are%
\begin{equation}
E_{nl\sigma}=E_{nl}+g\mu_{B}B\sigma, \label{Levels}%
\end{equation}
where $g$ is the Land\'{e} factor, $\mu_{B}$ is the Bohr magneton, $\sigma
=\pm1/2$ is the spin projection. The energy levels $E_{nl}$ corresponding to
the wave functions $\Psi_{nlm}\left(  \rho,\varphi,z\right)  $ in
(\ref{States}) are calculated using the five-band $\mathbf{k\cdot p}$ model
\cite{Pfeffer1996}.

\subsection{Optical conductivity}

Within the memory-function formalism, the real part of the frequency-dependent
conductivity (in the Faraday configuration) in the local parabolic band
approximation, can be written as \cite{DW87,Wu1989}%
\begin{equation}
\operatorname{Re}\sigma\left(  \omega\right)  =-\frac{n_{S}e^{2}}{m_{b}%
}\operatorname{Im}\frac{1}{\omega-\omega_{c}-\chi\left(  \omega,\omega
_{c}\right)  /\omega+i\delta}\quad\left(  \delta\rightarrow+0\right)  ,
\label{ReSigma}%
\end{equation}
where $n_{S}$ is the 2D electron density, and $\chi\left(  \omega\right)  $ is
the memory function \cite{DW87}. To take into account the splitting of the CR
peaks due to the \emph{nonparabolicity} of the conduction band, Eq.
(\ref{ReSigma}) is generalized as follows:%
\begin{equation}
\operatorname{Re}\sigma\left(  \omega\right)  =\sum_{n,l,\sigma}%
\varkappa_{nl\sigma}\operatorname{Re}\sigma_{nl\sigma}\left(  \omega\right)  ,
\label{Resigma1}%
\end{equation}
where each contribution $\operatorname{Re}\sigma_{nl\sigma}\left(
\omega\right)  $, corresponding to the transitions $\left(  l\rightarrow
l+1\right)  $,%
\begin{equation}
\operatorname{Re}\sigma_{nl\sigma}\left(  \omega\right)  =-\frac{n_{S}e^{2}%
}{m_{b}}\operatorname{Im}\frac{1}{\omega-\omega_{c}^{\left(  nl\sigma\right)
}-\chi\left(  \omega,\omega_{c}^{\left(  nl\sigma\right)  }\right)
/\omega+i\delta}\quad\left(  \delta\rightarrow+0\right)  \label{REsigma2}%
\end{equation}
is calculated with the local parabolic band approximation. The transition
frequency%
\begin{equation}
\omega_{c}^{\left(  nl\sigma\right)  }\equiv\frac{E_{n,l+1,\sigma}%
-E_{nl\sigma}}{\hbar} \label{wc}%
\end{equation}
is used in Eq. (\ref{REsigma2}) as distinct from the cyclotron frequency
$\omega_{c}$ in Eq. (\ref{ReSigma}). The weight $\varkappa_{nl\sigma}$ is
proportional to the number of open channels for the transitions $\left(
l\rightarrow l+1\right)  $. The normalized weights $\varkappa_{nl\sigma}$ are%
\begin{equation}
\varkappa_{nl\sigma}=\frac{f_{nl\sigma}\left(  1-f_{n,l+1,\sigma}\right)
}{\sum_{n^{\prime},l^{\prime},\sigma^{\prime}}f_{n^{\prime}l^{\prime}%
\sigma^{\prime}}\left(  1-f_{n^{\prime},l^{\prime}+1,\sigma^{\prime}}\right)
}, \label{weight}%
\end{equation}
where $f_{nl\sigma}$ is the Fermi occupation number. The memory function
$\chi\left(  \omega,\omega_{c}\right)  $ takes the form (cf.
Refs.~\cite{DW87,HPD93})%
\begin{align}
\chi\left(  \omega,\omega_{c}\right)   &  =-\sum_{n,k}\sum_{\lambda}%
\int\frac{d^{2}\mathbf{q}}{\left(  2\pi\right)  ^{2}}\frac{\left|  \left(
\gamma_{\lambda,\mathbf{q}}\right)  _{nk}\right|  ^{2}q^{2}}{n_{S}\hbar
m_{b}\epsilon^{2}\left(  q\right)  }\nonumber\\
&  \times\int\limits_{0}^{\infty}dte^{-\delta t}\left(  e^{i\omega
t}-1\right)  \operatorname{Im}\left[  \mathcal{D}_{_{\lambda}}\left(
\mathbf{q,}t\right)  G_{nk}\left(  \mathbf{q},t\right)  \right]  \quad\left(
\delta\rightarrow+0\right)  , \label{MemFun}%
\end{align}
where $\epsilon\left(  q\right)  $ is the static screening factor \cite{DW87}.
$\mathcal{D}_{_{\lambda}}\left(  \mathbf{q,}t\right)  $ is the phonon Green's
function%
\begin{equation}
\mathcal{D}_{_{\lambda}}\left(  \mathbf{q,}t\right)  \equiv-i\Theta\left(
t\right)  \left[  \left\langle b_{\lambda,\mathbf{q}}\left(  t\right)
b_{\lambda,\mathbf{q}}^{+}\left(  0\right)  \right\rangle +\left\langle
b_{\lambda,-\mathbf{q}}^{+}\left(  t\right)  b_{\lambda,-\mathbf{q}}\left(
0\right)  \right\rangle \right]  \label{FP}%
\end{equation}
and $G_{nk}\left(  \mathbf{q},t\right)  $ is the electron density-density
Green's function%
\begin{equation}
G_{nk}\left(  \mathbf{q},t\right)  =-i\Theta\left(  t\right)  \frac{1}%
{S}\left\langle \rho_{nk}\left(  \mathbf{q},t\right)  \rho_{nk}^{+}\left(
\mathbf{q},0\right)  \right\rangle \label{GF}%
\end{equation}
with the Heaviside step function $\Theta\left(  t\right)  $. The averaging in
Eqs. (\ref{FP}) and (\ref{GF}) is performed on the equilibrium statistical
operator of the electron-phonon system.

In the present paper, we apply the method presented above to the case of a
GaAs/AlAs quantum well, where the weak-coupling regime is realized. To obtain
$\chi\left(  \omega,\omega_{c}\right)  $ to second order in perturbation
theory, the electron density-density Green's function $G_{nk}\left(
\mathbf{q},t\right)  $ can be calculated neglecting the electron-phonon
interaction. The electron-electron interaction is taken into account in the
random-phase approximation (RPA) following Ref.~\cite{Backes1992}, where
plasmon modes are derived for electrons in a layer. As a result, we arrive at
the following set of equations in the Fourier representation%
\begin{equation}
\sum_{n_{1}k_{1}}\left[  \delta_{nn_{1}}\delta_{kk_{1}}-P_{nk}\left(
\mathbf{q},\omega\right)  V_{C}\left(  k,n;k_{1},n_{1}|\mathbf{q}\right)
\right]  G_{n_{1}k_{1},\tilde{n}\tilde{k}}^{R}\left(  \mathbf{q}%
,\omega\right)  =\delta_{n\tilde{n}}\delta_{k\tilde{k}}\hbar P_{nk}\left(
\mathbf{q},\omega\right)  . \label{Set}%
\end{equation}
Here, the auxiliary retarded Green's function $G_{nk,\tilde{n}\tilde{k}}%
^{R}\left(  \mathbf{q},\omega\right)  $ is determined as%
\begin{equation}
G_{nk,\tilde{n}\tilde{k}}^{R}\left(  \mathbf{q},\omega\right)  \equiv
-\frac{i}{S}\int_{0}^{\infty}\left\langle \left[  \rho_{nk}\left(
\mathbf{q},t\right)  ,\rho_{\tilde{n}\tilde{k}}^{+}\left(  \mathbf{q}%
,0\right)  \right]  \right\rangle e^{i\omega t-\delta t}dt\quad\left(
\delta\rightarrow+0\right)  . \label{Auxil}%
\end{equation}
The RPA structure factor $P_{nk}\left(  \mathbf{q},\omega\right)  $ is
proportional to the retarded density-density Green's function of
non-interacting electrons in the quantum well in the presence of a magnetic
field. It is calculated explicitly, using the second-quantization
representation of the density operators (\ref{density}), and takes the form%
\begin{align}
P_{nk}\left(  \mathbf{q},\omega\right)   &  =-i\frac{m_{b}\omega_{c}}%
{2\pi\hbar^{2}}\int_{0}^{\infty}dt\exp\left[  i\left(  \omega-\omega
_{kn}\right)  t-\delta t\right]  \sum_{l\sigma}L_{l}^{\left(  0\right)
}\left(  \frac{\hbar q^{2}}{m_{b}\omega_{c}}\left(  1-\cos\omega_{c}t\right)
\right) \nonumber\\
&  \times\left(  f_{nl\sigma}\exp\left[  -\frac{\hbar q^{2}}{2m_{b}\omega_{c}%
}\left(  1-e^{-i\omega_{c}t}\right)  \right]  -f_{kl\sigma}\exp\left[
-\frac{\hbar q^{2}}{2m_{b}\omega_{c}}\left(  1-e^{i\omega_{c}t}\right)
\right]  \right) \label{P}\\
&  \left(  \delta\rightarrow+0\right)  ,\nonumber
\end{align}
where $\omega_{kn}$ is the transition frequency between the $k$-th and the
$n$-th size-quantized subbands.

The set of equations (\ref{Set}) is valid for arbitrary electron density. The
Green's function $G_{nk}\left(  \mathbf{q},\omega\right)  $ is obtained from
$G_{nk}^{R}\left(  \mathbf{q},\omega\right)  $ using the Kramers-Kronig
dispersion relations and the relation, which follows from the analytical
properties of Green's functions,%
\begin{equation}
\operatorname{Im}G_{nk}^{R}\left(  \mathbf{q},\omega\right)  =\left(
1-e^{-\beta\omega}\right)  \operatorname{Im}G_{nk}\left(  \mathbf{q}%
,\omega\right)  , \label{R1}%
\end{equation}
where $\beta\equiv1/\left(  k_{B}T\right)  $. Further on, we consider the case
of a sufficiently narrow quantum well, when only the lowest size-quantization
subband $\left(  n=1\right)  $ is filled, and the transition frequency
$\omega_{21}\gg\omega_{Li},$ where $\omega_{Li}$ is the LO-phonon frequency in
the $i$-th medium ($i=1,2$). Under these conditions, transitions of an
electron to other subbands can be neglected. This results in the explicit
solution%
\begin{equation}
G_{11}^{R}\left(  \mathbf{q},\omega\right)  =\frac{\hbar P_{11}\left(
\mathbf{q},\omega\right)  }{1-V_{C}\left(  \mathbf{q}\right)  P_{11}\left(
\mathbf{q},\omega\right)  } \label{expl}%
\end{equation}
with $V_{C}\left(  \mathbf{q}\right)  \equiv V_{C}\left(  1,1;1,1|\mathbf{q}%
\right)  .$

The phonon retarded Green's function is calculated within the random-phase
approximation taking into account the magneto-plasmon-phonon mixing (cf.
Ref.~\cite{Wendler90}). In the present case of a narrow quantum well, we take
into account the mixing of phonons with intrasubband magneto-plasmons only. In
this approximation, disentangling a set of coupled Dyson equations leads to:%
\begin{equation}
\mathcal{D}_{\lambda}^{R}\left(  \mathbf{q},\omega\right)  =\frac{2\left(
\omega_{\lambda,q}+\frac{\left|  \left(  \gamma_{\lambda,q}\right)
_{11}\right|  ^{2}}{\hbar^{2}}G_{11}^{R}\left(  \mathbf{q},\omega\right)
\right)  }{\left(  \omega+i\delta\right)  ^{2}-\omega_{\lambda,q}^{2}%
-2\omega_{\lambda,q}\frac{\left|  \left(  \gamma_{\lambda,q}\right)
_{11}\right|  ^{2}}{\hbar^{2}}G_{11}^{R}\left(  \mathbf{q},\omega\right)
}\quad\left(  \delta\rightarrow+0\right)  . \label{qq9}%
\end{equation}
The poles of this Green's function are the roots of the equation
\begin{equation}
\omega^{2}-\omega_{\lambda,q}^{2}-2\omega_{\lambda,q}\frac{\left|  \left(
\gamma_{\lambda,q}\right)  _{11}\right|  ^{2}}{\hbar^{2}}\operatorname{Re}%
G_{11}^{R}\left(  \mathbf{q},\omega\right)  =0 \label{Poles}%
\end{equation}
and they determine the spectrum of mixed magneto-plasmon-phonon excitations in
the quantum well. Let $\Omega_{\lambda,j}\left(  \mathbf{q}\right)  $ denote
the positive roots of Eq. (\ref{Poles}). Using $G_{11}^{R}\left(
\mathbf{q},\omega\right)  $ given by Eq. (\ref{expl}), with the structure
factor (\ref{P}), we find that the retarded Green's function (\ref{qq9}) can
be expanded as a series%
\begin{equation}
\mathcal{D}_{\lambda}^{R}\left(  \mathbf{q},\omega\right)  =\sum_{j}%
A_{\lambda,j}\left(  \mathbf{q}\right)  \left(  \frac{1}{\omega-\Omega
_{\lambda,j}\left(  \mathbf{q}\right)  +i\delta}-\frac{1}{\omega
+\Omega_{\lambda,j}\left(  \mathbf{q}\right)  +i\delta}\right)  ,\quad
\delta\rightarrow+0, \label{Exp1}%
\end{equation}
where $A_{\lambda,j}\left(  \mathbf{q}\right)  $ is the residue of
$\mathcal{D}_{\lambda}^{R}\left(  \mathbf{q},\omega\right)  $ at
$\omega=\Omega_{\lambda,j}\left(  \mathbf{q}\right)  .$ The Green's function
$\mathcal{D}_{\lambda}\left(  \mathbf{q},\omega\right)  $ is obtained from
$\mathcal{D}_{\lambda}^{R}\left(  \mathbf{q},\omega\right)  $ using relations
between $\operatorname{Im}\mathcal{D}_{\lambda}^{R}\left(  \mathbf{q}%
,\omega\right)  $ and $\operatorname{Im}\mathcal{D}_{\lambda}\left(
\mathbf{q},\omega\right)  $ similar to Eq. (\ref{R1}), and the Kramers-Kronig
dispersion relations. As a result, the Green's function $\mathcal{D}_{\lambda
}\left(  \mathbf{q},\omega\right)  $ can be written down explicitly,%
\begin{equation}
\mathcal{D}_{\lambda}\left(  \mathbf{q},\omega\right)  =\sum_{j}%
\frac{A_{\lambda,j}\left(  \mathbf{q}\right)  }{1-e^{-\beta\Omega_{\lambda
,j}\left(  \mathbf{q}\right)  }}\left(  \frac{1}{\omega-\Omega_{\lambda
,j}\left(  \mathbf{q}\right)  +i\delta}+\frac{e^{-\beta\Omega_{\lambda
,j}\left(  \mathbf{q}\right)  }}{\omega+\Omega_{\lambda,j}\left(
\mathbf{q}\right)  +i\delta}\right)  ,\quad\delta\rightarrow+0. \label{Exp2}%
\end{equation}
It is seen from Eqs. (\ref{MemFun}) and (\ref{Exp2}), that the parameter%
\begin{equation}
w_{\lambda,j}\left(  \mathbf{q}\right)  \equiv\left|  \left(  \gamma
_{\lambda,\mathbf{q}}\right)  _{11}\right|  ^{2}A_{\lambda,j}\left(
\mathbf{q}\right)  \label{Cstr}%
\end{equation}
determines the strength of the coupling between an electron and the $\left(
\lambda,j\right)  $-th magneto-plasmon-phonon mode characterized by the wave
vector $\mathbf{q}$.

\section{Results and discussion}

Using the method developed in the previous section, the magneto-plasmon-phonon
frequencies and the magneto-absorption spectra of a GaAs/AlAs quantum well are
now calculated numerically. Within this approach, the CR peaks are
proportional to delta-functions because of the discrete energy spectrum of an
electron in a quantum well in the presence of a magnetic field (see, e.~g.,
Ref.~\cite{HPD93}). In what follows, we introduce in Eq. (\ref{MemFun}) a
finite broadening $\gamma$ of the Landau levels instead of the infinitesimal
parameter $\delta\rightarrow+0$. Here, we choose the value $\gamma
=0.01\omega_{L1}$, which corresponds to the linewidth of the CR peaks in the
experiment of Ref.~\cite{Poulter2001}. The material parameters used for the
present calculation are listed in Table 1.

\bigskip

\begin{center}
\textbf{Table 1. Material parameters used for the calculation of the
magneto-absorption spectra of a GaAs/AlAs quantum well}

\medskip%

\begin{tabular}
[c]{|c|c|c|}\hline
Parameter & GaAs & AlAs\\\hline
LO-phonon energy, meV & 36.3 \cite{Poulter2001} & 50.09 \cite{Adachi}\\\hline
TO-phonon energy, meV & 33.6 \cite{Poulter2001} & 44.88 \cite{Adachi}\\\hline%
\begin{tabular}
[c]{l}%
Electron band mass, in units\\
of the electron mass in the vacuum
\end{tabular}
& 0.0653 \cite{Pfeffer1996} & 0.15 \cite{Adachi}\\\hline
High-frequency dielectric constant & 10.89 \cite{Adachi} & 8.16 \cite{Adachi}%
\\\hline
Electron-phonon coupling constant & 0.068 \cite{D1972} & 0.148 \cite{Adachi}%
\\\hline
\end{tabular}

\bigskip
\end{center}

\noindent The value 0.915 eV \cite{HPD93} is used for the AlAs/GaAs potential
barrier. In the numerical calculation, we use a formula for the energy levels
in a non-parabolic conduction band taken from Refs.~\cite{Wu1989,Pfeffer1996}
\begin{equation}
E_{nl}=\frac{E_{0}^{\ast}}{2}\left(  \sqrt{1+4\frac{E_{nl}^{0}}{E_{0}^{\ast}}%
}-1\right)  \label{np3}%
\end{equation}
with the effective energy gap $E_{0}^{\ast}=0.98$ eV \cite{Pfeffer1996} and
with $E_{nl}^{0}=\hbar\omega_{c}\left(  l+\frac{1}{2}\right)  +\varepsilon
_{n}$, where $\varepsilon_{n}$ is a size-quantized energy level characterizing
the motion along the $z$-axis.

According to Ref.~\cite{HPD93}, the following phonon branches give
contributions to the CR spectra in a quantum well: (i)~\emph{bulk-like
phonons} with frequency $\omega_{L1}$, (ii) \emph{symmetric interface phonons}
with frequencies $\omega_{S,\pm}\left(  q\right)  $ given by Eq. (7) of
Ref.~\cite{HPD93}. In a similar way, the magneto-plasmon-phonon modes in a
quantum well can be classified into bulk-like and interface modes. The
frequencies of these modes depend on the electron density and on the modulus
$q$ of the 2D wave vector $\mathbf{q}$. As follows from our calculations, the
modes, whose wave vectors lie in the region with $q\sim2q_{p},$ where
$q_{p}\equiv\left(  m_{b}\omega_{L1}/\hbar\right)  ^{1/2},$ give the dominant
contribution to the magneto-optical absorption. In Fig. 1, frequencies of six
mixed magneto-plasmon-phonon modes, which contribute significantly to the
cyclotron-resonance spectrum, are plotted as a function of the electron
density for $\omega_{c}=0.8\omega_{L1}$ and for $q=2q_{p}$. In the low-density
limit, as seen from Fig.~1, the phonon modes do not mix with the
magneto-plasmons. At $n_{S}\rightarrow0,$ the frequencies of the bulk-like and
of two symmetric interface magneto-plasmon-phonon modes tend to $\omega_{L1}$
and $\omega_{S,\pm}\left(  q\right)  $, respectively (cf. Ref.~\cite{HPD93}).
With increasing density $n_{S},$ the magneto-plasmon branch splits into three
modes (shown by black curves). When the density rises further, anticrossing of
phonon and magneto-plasmon frequencies is to be expected between $10^{11}%
$~cm$^{-2}\lesssim n_{S}\lesssim10^{12}$~cm$^{-2}$. This anticrossing allows
us to distinguish between the lower-frequency modes (black curves) and the
higher-frequency modes (red curves). Finally, for $n_{S}\gtrsim10^{12}$
cm$^{-2}$, the three higher frequencies tend to the magneto-plasmon frequency,
which behaves as $\sim n_{S}^{1/2},$ and the three lower frequencies are close
to $\omega_{T1}$ and to $\omega_{S,\pm}\left(  q\right)  $, respectively. The
latter two magneto-plasmon-phonon frequencies are actually slightly reduced
with respect to $\omega_{S,\pm}\left(  q\right)  $. The appearance of the
branch with frequency close to $\omega_{T1}$ instead of $\omega_{L1}$ is due
to the screening of the electron-phonon interaction by the plasma vibrations.

In Fig. 2, the relative contributions of different magneto-plasmon-phonon
modes to the total magneto-polaron coupling strength
\begin{equation}
\tilde{w}_{\lambda,j}\left(  q\right)  \equiv\frac{w_{\lambda,j}\left(
q\right)  }{\sum_{\lambda,j}w_{\lambda,j}\left(  q\right)  } \label{rcw}%
\end{equation}
are plotted as a function of $n_{S}$ for $q=2q_{p}.$ In the low-density limit,
the plasma vibrations give a vanishing contribution to $\tilde{w}_{\lambda
,j}\left(  q\right)  $. In this case, the electrons interact only with
phonons. For increasing electron density, the strength of the interaction of
an electron with the lower-frequency magneto-plasmon-phonon modes rises, while
that for the higher-frequency modes decreases. For a 10-nm GaAs/AlAs quantum
well, the contribution of the interface magneto-plasmon-phonon modes into the
CR spectra is smaller than that of the bulk-like modes. Nevertheless, the
interaction of an electron with the interface modes is not negligible.

It was noted in Refs.~\cite{Kallin85,Zhang02}, that the participation of a
magneto-plasmon at $\mathbf{q}\neq0$ in CR for a purely electronic system in a
quantum well, without impurities, is not allowed by momentum conservation.
However, for a polaron gas in a quantum well, the interaction of the electrons
with the magneto-plasmon-phonon modes characterized in Fig. 2 is reflected in
the CR spectrum. It is remarkable that, in the high-density limit, the largest
coupling strength occurs for the mixed magneto-plasmon-phonon mode with
$\omega\approx\omega_{T1}.$ This fact provides a quantitative basis for the
understanding of the resonant magneto-polaron coupling observed in
Ref.~\cite{Poulter2001}. A related effect appears also for the magneto-phonon
resonance: as shown in Ref.~\cite{Afonin2000}, the magneto-plasmon-phonon
mixing leads to a shift of the resonant frequency of the magneto-phonon
resonance in quantum wells from $\omega\approx\omega_{L1}$ to $\omega
\approx\omega_{T1}$.

In Ref.~\cite{Poulter2001}, CR spectra are measured for the absolute
transmission $T\left(  \omega\right)  ,$ which is related to the optical
absorption coefficient by%
\begin{equation}
T\left(  \omega\right)  \equiv1-\Gamma\left(  \omega\right)  . \label{tr}%
\end{equation}
The optical absorption coefficient is proportional to $\operatorname{Re}%
\sigma\left(  \omega\right)  $,%
\begin{equation}
\Gamma\left(  \omega\right)  =\frac{4\pi}{cn\left(  \omega\right)
}\operatorname{Re}\sigma\left(  \omega\right)  , \label{abs}%
\end{equation}
where $n\left(  \omega\right)  $ is the refractive index. In Fig. 3, we
compare the experimental \cite{Poulter2001} transmission spectra with the
transmission spectra calculated in the present work for a polaron gas with
electron density $n_{S}=1.28\times10^{12}$ cm$^{-2}$ in a 10-nm GaAs/AlAs
quantum well. Reliable comparison of our theoretical results with the
experimental data of Ref.~\cite{Poulter2001} can be performed only outside a
frequency band near $\omega_{T1}$ (shown in Fig. 3 by the cross-hatched area),
because the experimental spectra in this frequency region are obtained after
the elimination of a strong absorption peak due to direct TO-phonon absorption
\emph{in the substrate}.\ It is not straightforward to extract information on
the CR spectra in this frequency region. It is seen from Fig. 3, that outside
the cross-hatched area the theoretical peak positions of the
magneto-transmission spectra calculated in the present work reasonably compare
with the experimental data of Ref.~\cite{Poulter2001}. If one tunes the
magnetic field within the error bar, the agreement between theory and
experiment can be improved. The remaining difference between theoretical and
experimental peak positions can be explained by possible deviation of the
values of the band parameters (the electron band mass and the effective energy
gap) in a GaAs/AlAs quantum well from those in bulk.

The effect of the magneto-plasmon-phonon mixing is clearly manifested in the
CR spectra, as seen, e. g., from the plot corresponding to a magnetic field
$B=22.25$ T, for which $\omega_{c}\approx\omega_{T1}\approx33.6$ meV. For this
value of $B$, the intensities of the peaks below $\omega_{T1}$ and above
$\omega_{T1}$ are comparable both in the experiment \cite{Poulter2001} and in
the present theory. Such a behavior is typical for resonant magneto-polaron
coupling near $\omega_{T1}$ rather than near $\omega_{L1}$.

The splitting of the CR peaks manifested in Fig. 3 reflects the
nonparabolicity of the conduction band. In the experimental CR spectra
\cite{Poulter2001}, the higher-frequency and lower-frequency components of the
split peaks are denoted as A- resp. B-components. The higher-frequency
component of each split peak corresponds to the transitions between the Landau
levels $\left(  0\rightarrow1\right)  $, while the lower-frequency component
corresponds to the transitions $\left(  1\rightarrow2\right)  $. With
increasing magnetic field strength, the distance between Landau levels rises,
so that the upper filled level ($l=1$) becomes less populated. As a result,
the relative intensity of the lower-frequency peak diminishes. It is shown in
Ref.~\cite{Manger2001}, that the observed relative oscillator strengths of the
split CR peaks are not consistent with the one-particle theoretical picture.
The one-particle theories (e. g., Refs.~\cite{Kobayashi74,Ando75})
overestimate the lower-frequency peak intensity as compared to experiment. In
particular, for a filling factor $\nu=3,$ the one-particle approximation
predicts $I_{\left(  1\rightarrow2\right)  }/I_{\left(  0\rightarrow1\right)
}\approx2$, which contradicts the experimental CR data for high electron
densities \cite{Manger2001}, where this ratio is close to 1. It is noteworthy,
that for $\nu=3,$ the coefficients $\varkappa_{nl\sigma}$ given by
(\ref{weight}) for $l=0$ and for $l=1$ are equal to each other, while the
contributions $\operatorname{Re}\sigma_{nl\sigma}\left(  \omega\right)  $ only
slightly depend on the Landau level quantum number. Therefore, for $\nu=3,$
our theory gives $I_{\left(  1\rightarrow2\right)  }/I_{\left(  0\rightarrow
1\right)  }\approx1$, in agreement with Ref. \cite{Manger2001}.

In Fig. 4, we plot a set of our theoretical magneto-absorption spectra for a
GaAs/AlAs quantum well as a color map. This map shows the magneto-absorption
intensity as a function of both the magnetic field and the frequency. For
comparison, the experimental peak position taken from Fig. 3 of
Ref.~\cite{Poulter2001} are shown in the same graph. For the peaks with
$B\lessapprox17.2$ ~T (indicated by crosses) the splitting due to the band
nonparabolicity is not resolved experimentally. It is seen from Fig. 4 that,
for the high-density electron gas, anticrossing of the CR spectra occurs near
$\omega_{T1}$ both for the experimental and for the calculated CR spectra.
Starting from large $B,$ a set of peaks with $\omega>\omega_{T1}$ are pinned
towards $\omega_{T1}$ from above, while starting from $B\sim0,$ another set of
peaks are pinned from below to a frequency slightly larger than $\omega_{T1}.$

For a low-density polaron gas in a GaAs/AlAs quantum well, it was found
\cite{HPD93}, that the resonance frequencies, calculated taking into account
the interaction of electrons with both bulk-like and interface phonons,
appreciably deviate from those found for the interaction with bulk LO phonons
only. In particular, due to interface phonons, a resonance frequency appears
between $\omega_{L1}$ and $\omega_{T1}$ (see Fig. 7 of Ref. \cite{HPD93}). In
the present work, we take into account, in addition to the effects due to the
interaction of electrons with bulk-like and interface phonons, the many-body
effects, which ensure the shift of the resonance frequency towards
$\omega_{T1},$ when the electron density is high enough.

The splitting of the magneto-absorption peaks due to the band nonparabolicity,
observed in the experiment for $\omega<\omega_{T1},$ continues in the
theoretically calculated spectra also in the region $\omega>\omega_{T1}$.
However, the intensity of the B-component for $\omega>\omega_{T1}$ is quite small.

\section{Conclusions}

We have theoretically investigated cyclotron-resonance spectra for a polaron
gas in a GaAs/AlAs quantum well with high electron density, taking into
account (i) the electron-electron interaction and as a consequence the
screening of the electron-phonon interaction, (ii) the magneto-plasmon-phonon
mixing, (iii) the electron-phonon interaction with relevant phonon modes
specific for the quantum well under consideration. As a result of the mixing,
different magneto-plasmon-phonon modes appear in the quantum well and
contribute to the CR spectra. In the 10-nm GaAs/AlAs quantum well
(investigated experimentally in Ref.~\cite{Poulter2001}), the interaction of
an electron with the bulk-like modes is dominant. It is shown in the present
work that, in the high-density polaron gas, the resonant magneto-polaron
coupling, manifested in the CR spectra, takes place at a frequency close to
the TO-phonon frequency in GaAs. This effect is in contrast with the cyclotron
resonance of a low-density polaron gas in a quantum well, where anticrossing
occurs near the LO-phonon frequency. Due to the band nonparabolicity, the CR
spectrum of a high-density polaron gas in a quantum well is split into two
components. The calculated CR-spectra are in agreement with experiment, in
particular, with the results of Ref.~\cite{Poulter2001}.

\begin{acknowledgments}
Discussions with V. M. Fomin, G. Martinez and V. N. Gladilin are gratefully
acknowledged. This work has been supported by the GOA BOF UA 2000, I.U.A.P.,
F.W.O.-V. projects G.0274.01N, G.0435.03, the W.O.G. WO.025.99 (Belgium) and
the European Commission GROWTH Programme, NANOMAT project, contract No. G5RD-CT-2001-00545.
\end{acknowledgments}

\newpage

\begin{quote}
\textbf{Figure captions}
\end{quote}

\bigskip

Fig. 1 (color). Frequencies of magneto-plasmon-phonon modes in a 10-nm
GaAs/AlAs quantum well as a function of the electron density. The value
$\omega_{c}=0.8\omega_{L1}$ is taken for the cyclotron frequency. Solid,
dashed and dash-dotted curves correspond to bulk-like, interface (GaAs) and
interface (AlAs) modes, respectively.

\bigskip

Fig. 2 (color). Relative contributions of six magneto-plasmon-phonon modes to
the total magneto-polaron coupling strength $w\left(  q\right)  \equiv
\sum_{\lambda,j}w_{\lambda,j}\left(  q\right)  $ in a 10-nm GaAs/AlAs quantum
well as a function of the electron density. The other modes are neglected,
because their relative contribution is only about 10$^{-4}$ of $w\left(
q\right)  .$ The notations are the same as those of Fig. 1.

\bigskip

Fig. 3 (color). Experimental and theoretical transmission spectra, at
different magnetic fields, for a 10-nm GaAs/AlAs quantum well with electron
density $n_{S}=1.28\times10^{12}$ cm$^{-2}$. The black solid curves represent
the experimental data of Ref.~\cite{Poulter2001}. The red dashed curves
correspond to the present theory.

\bigskip

Fig. 4 (color). Color map of the cyclotron resonance spectra for a 10-nm
GaAs/AlAs quantum well. Symbols indicate the peak positions of the
experimental spectra (which are taken from Fig. 3 of Ref.~\cite{Poulter2001}).
The dashed lines show LO- and TO-phonon frequencies in GaAs.

\newpage%

\begin{figure}
[ptbh]
\begin{center}
\includegraphics[
height=7.459in,
width=5.9949in
]%
{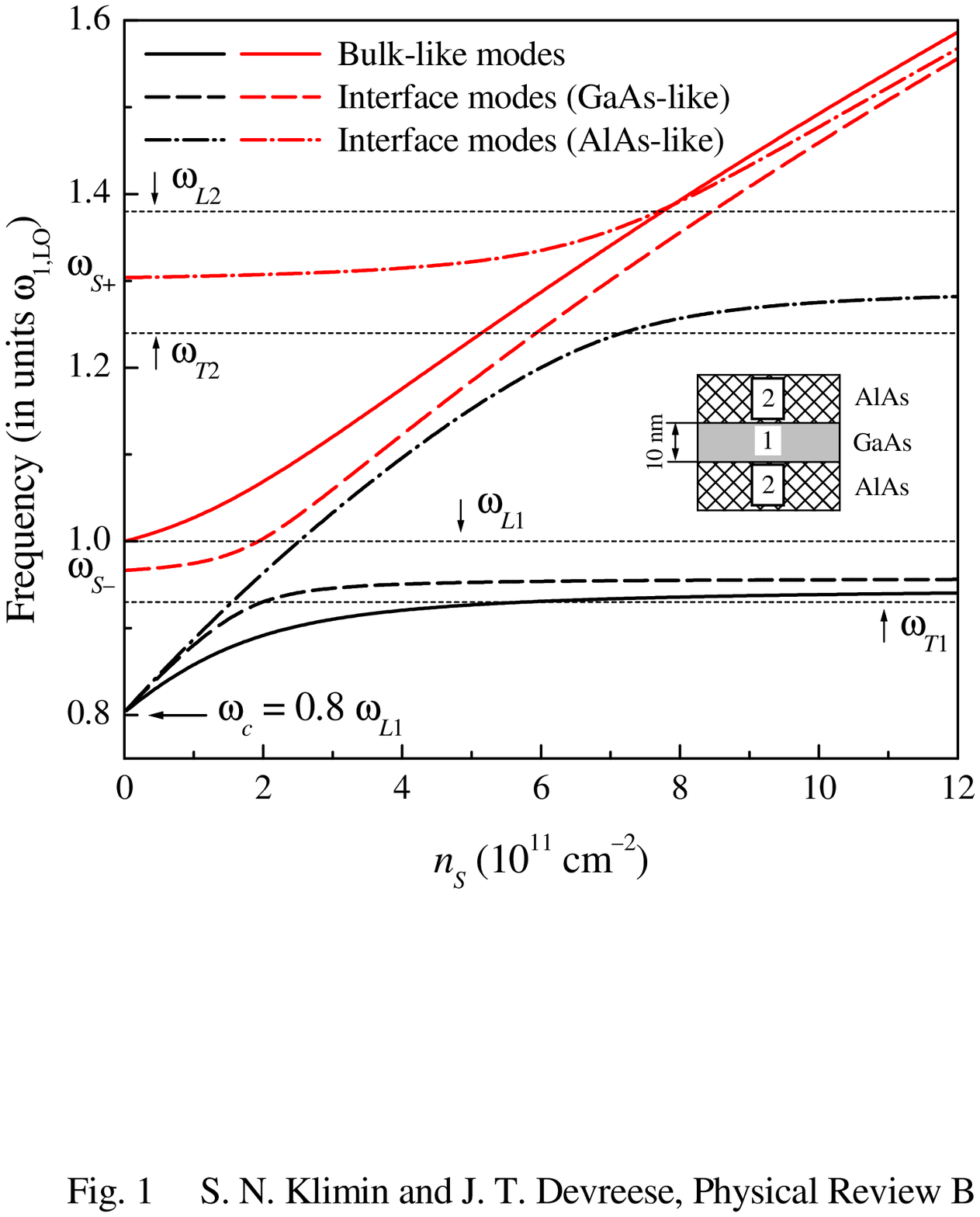}%
\end{center}
\end{figure}

\newpage%

\begin{figure}
[ptbh]
\begin{center}
\includegraphics[
height=7.3059in,
width=6.1782in
]%
{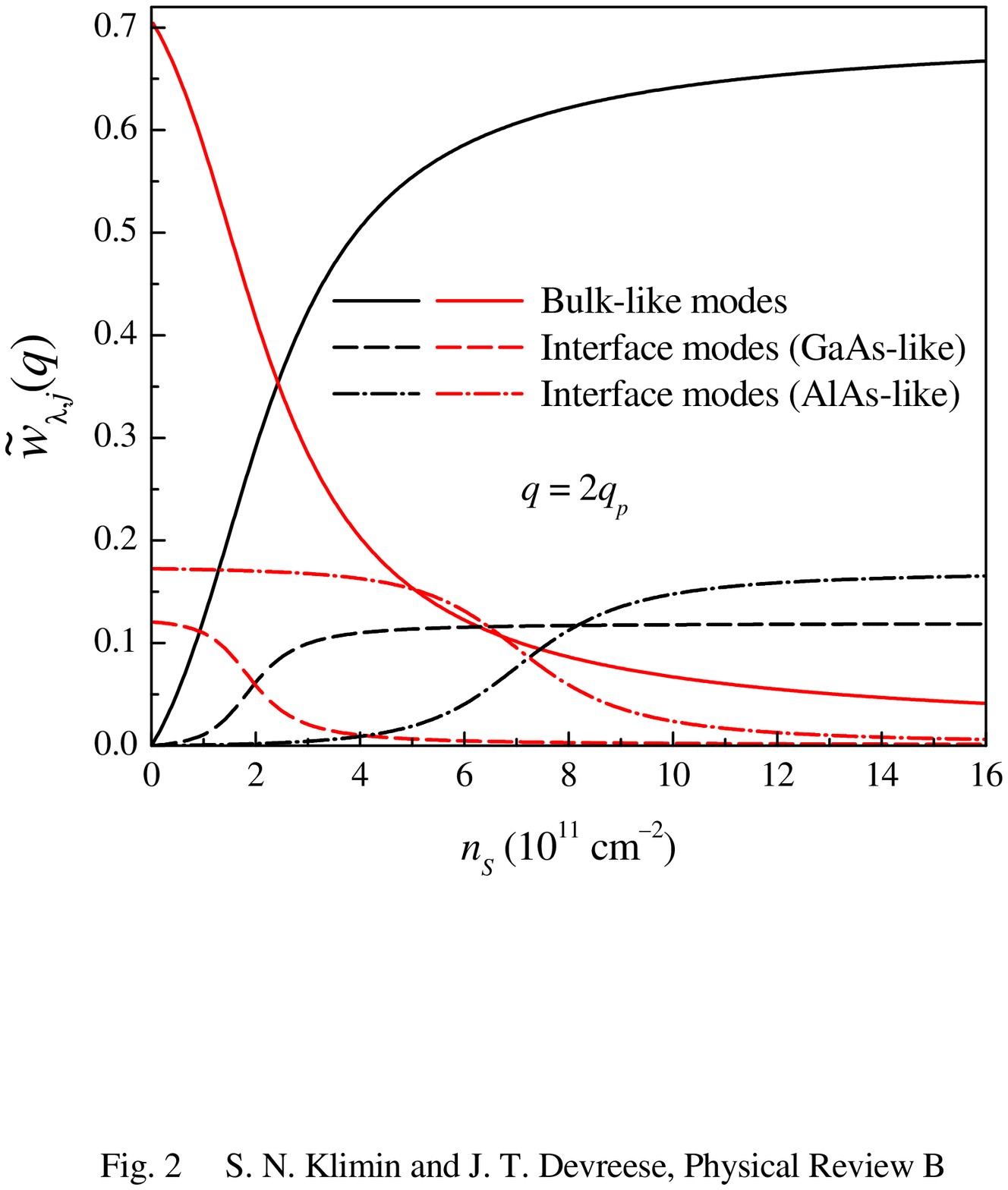}%
\end{center}
\end{figure}

\newpage%

\begin{figure}
[ptbh]
\begin{center}
\includegraphics[
height=7.67in,
width=4.4936in
]%
{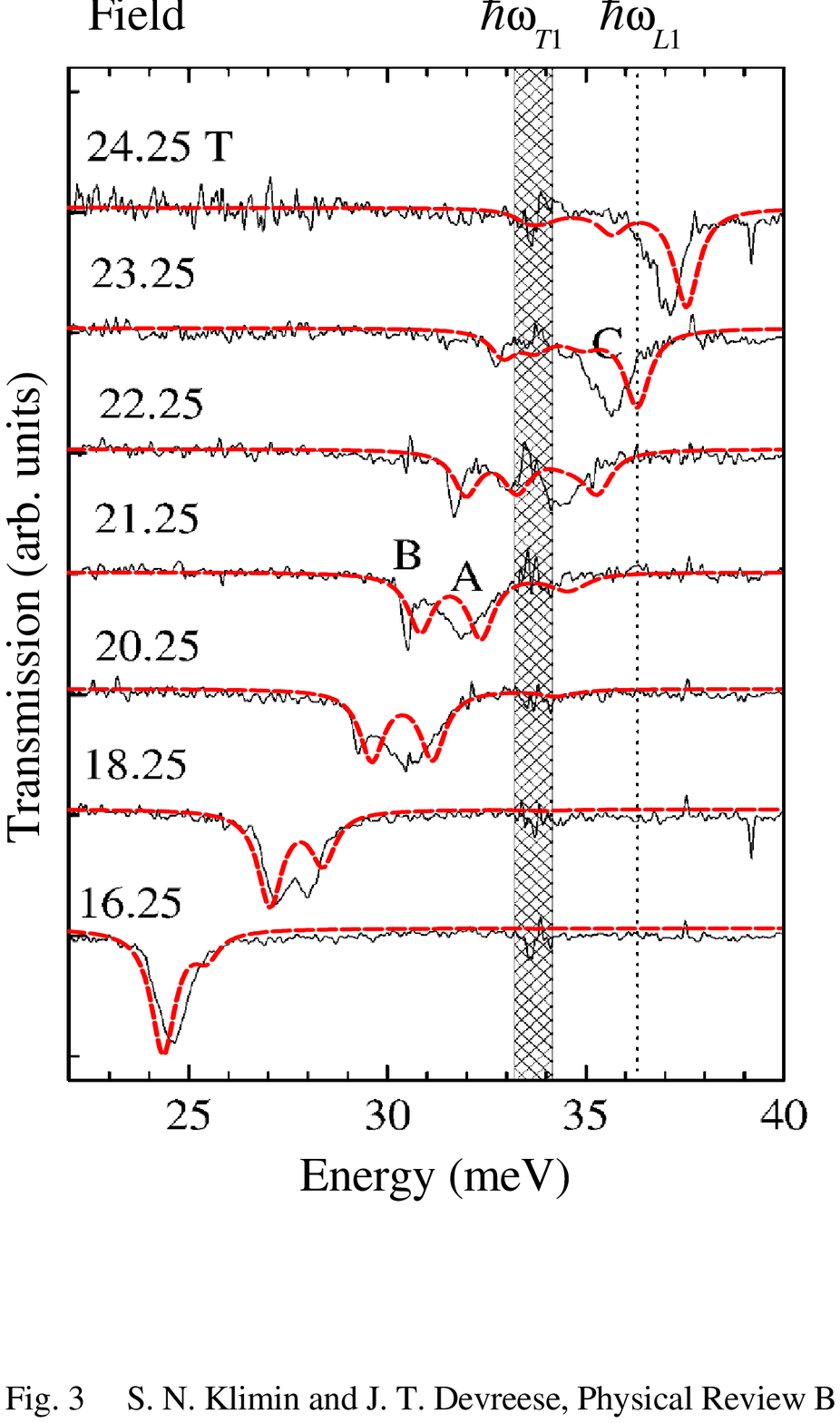}%
\end{center}
\end{figure}

\newpage%

\begin{figure}
[ptbh]
\begin{center}
\includegraphics[
height=5.7865in,
width=6.3667in
]%
{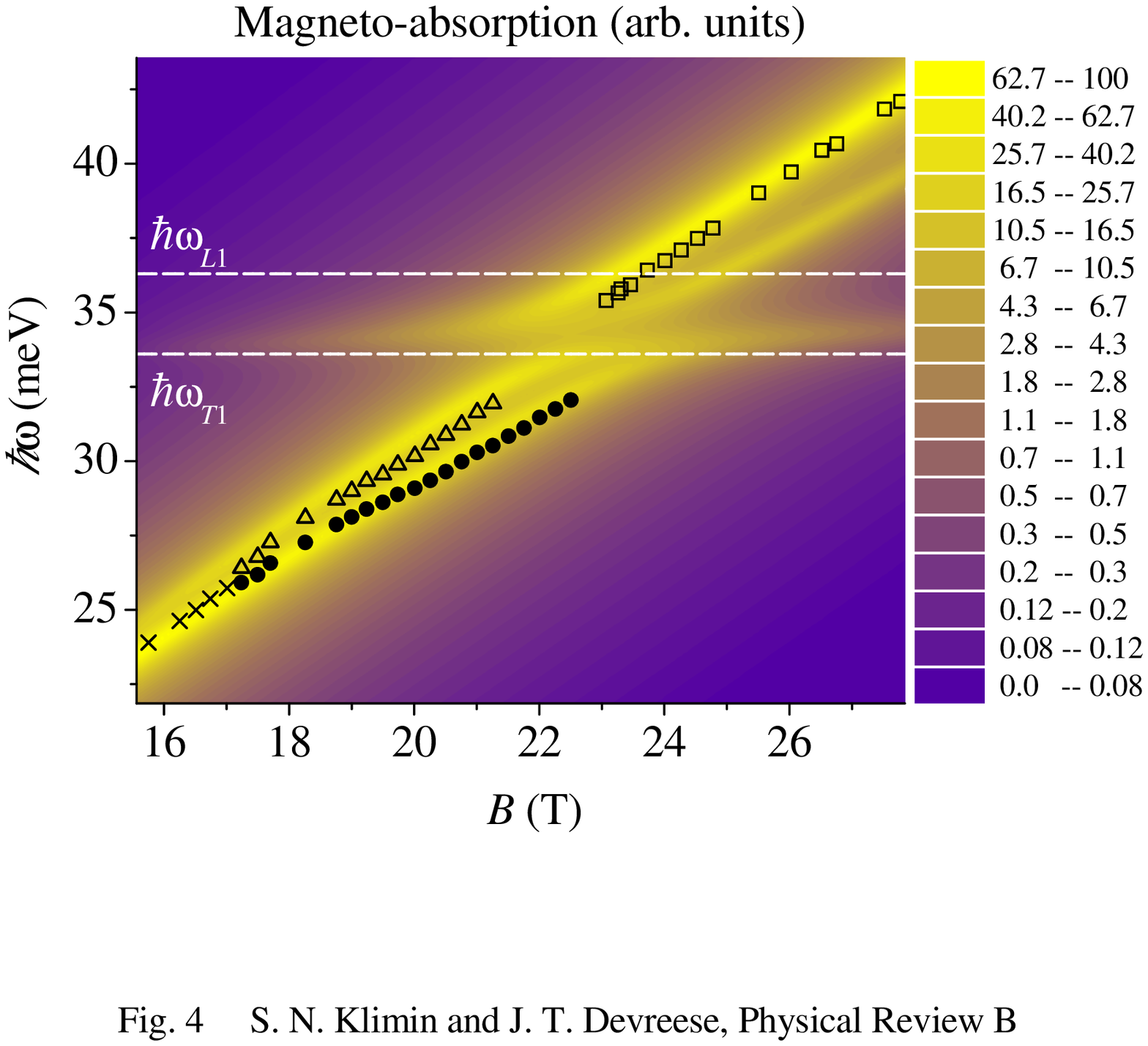}%
\end{center}
\end{figure}
\end{document}